\PassOptionsToPackage{table}{xcolor}

\documentclass[sigconf]{acmart}

\AtBeginDocument{%
  }

\renewcommand\footnotetextcopyrightpermission[1]{}

\usepackage{tikz}
\usepackage{graphicx}
\usepackage{amsmath}
\usepackage{subfigure}
\usepackage{stfloats}

\usepackage{epstopdf}
\usepackage{rotating}
\usepackage{cleveref}
\usepackage{multirow}
\usepackage{makecell}

\usepackage{siunitx}

\usepackage{pifont}

\makeatletter
\usepackage{comment}
\let\wfs@comment@comment\comment
\let\comment\@undefined

\usepackage{easyReview}
\let\wfs@easyReview@comment\comment
\let\comment\@undefined

\newcommand\comment{%
	\ifthenelse{\equal{\@currenvir}{comment}}
	{\wfs@comment@comment}
	{\wfs@easyReview@comment}%
}
\makeatother

\usepackage{algorithmicx}
\usepackage[ruled]{algorithm}
\usepackage[noend]{algpseudocode}
\usepackage{algpseudocode}

\usepackage{balance}

\usepackage{enumitem}

\graphicspath{{./}}

\crefformat{section}{\S#2#1#3} 
\crefformat{subsection}{\S#2#1#3}
\crefformat{subsubsection}{\S#2#1#3}

\crefname{section}{§}{§§}
\Crefname{section}{§}{§§}

\newcommand{\pname}[1]{CRouting{#1}}

\begin{document}
\begin{sloppypar}

\title[]{\pname{}: Reducing Expensive Distance Calls in Graph-Based Approximate Nearest Neighbor Search}

\author{Zhenxin Li}
\email{zhenxin@zju.edu.cn}
\affiliation{
  \institution{Zhejiang University}
  \city{Hangzhou}
  \country{China}
}

\author{Shuibing He}
\authornote{Shuibing He is the corresponding author.}
\email{heshuibing@zju.edu.cn}
\affiliation{
  \institution{Zhejiang University}
  \city{Hangzhou}
  \country{China}
}

\author{Jiahao Guo}
\email{guahao@zju.edu.cn}
\affiliation{
  \institution{Zhejiang University}
  \city{Hangzhou}
  \country{China}
}

\author{Xuechen Zhang}
\email{xuechen.zhang@wsu.edu}
\affiliation{
  \institution{Washington State University}
  \city{Vancouver}
  \country{United States of America}
}

\author{Xian-He Sun}
\email{sun@iit.edu}
\affiliation{
  \institution{Illinois Institute of Technology}
  \city{Chicago}
  \country{United States of America}
}

\author{Gang Chen}
\email{cg@zju.edu.cn}
\orcid{0000-0002-7483-0045}
\affiliation{
  \institution{Zhejiang University}
  \city{Hangzhou}
  \country{China}
}
\renewcommand{\shortauthors}{Li et al.}

\begin{abstract}

Approximate nearest neighbor search (ANNS) is a crucial problem in information retrieval and AI applications.
Recently, there has been a surge of interest in graph-based ANNS algorithms due to their superior efficiency and accuracy.
However, the repeated computation of distances in high-dimensional spaces constitutes the primary time cost of graph-based methods. 
To accelerate the search, we propose a novel routing strategy named \pname{}, which bypasses unnecessary distance computations by exploiting the angle distributions of high-dimensional vectors.
\pname{} is designed as a plugin to optimize existing graph-based search with minimal code modifications. 
Our experiments show that \pname{} reduces the number of distance computations by up to 41.5\% and boosts queries per second by up to 1.48$\times$ on two predominant graph indexes, HNSW and NSG.
Code is publicly available at \url{https://github.com/ISCS-ZJU/CRouting}.
\end{abstract}

\maketitle

\section{Introduction}
\label{sec:introduction}
Approximate nearest neighbor search (ANNS) finds wide application in areas such as information retrieval~\cite{information-bibm19,information-computer95}, image and video analysis~\cite{imi-tpami14}, 
key-value storage~\cite{tierbase-icde25},
and recommender systems~\cite{recommendation-www01},
where quick retrieval for the most similar $K$ items is crucial (known as the \textit{top-K} problem).
In particular, driven by recent advancements in large language models (LLMs), ANNS services have become a core component of modern AI infrastructure~\cite{rag-nips20}. Domain knowledge from various data formats (e.g., documents, images, and speech) is embedded and stored as high-dimensional feature vectors. When a user queries a chatbot, the ANNS engine retrieves semantically similar vectors, delivering relevant contexts to enhance the response quality of LLMs. 

The retrieval of exact nearest neighbors in high-dimensional spaces is computationally expensive due to the curse of dimensionality~\cite{curseofdim-98}.
To address this, ANNS offers a practical trade-off by returning an approximate set of the $K$ nearest neighbors within an acceptable error margin, thereby achieving efficient scalability for high-dimensional data.
Based on the index structure, the existing ANNS algorithms can be divided into four major categories, including tree-structure based approaches~\cite{kdtree-1975,covertree-icml06,mtree-vldb97,randomprojectiontree-stoc08}, hashing-based approaches~\cite{pstablelsh-scg04,dynamiclsh-sigmod12,qd-sigmod18,vhp-vldb20}, quantization-based approaches~\cite{aq-cvpr14,imi-tpami14,optpq-cvpr13,itq-tpami12}, and graph-based approaches~\cite{nssg-tpami21,nsg-vldb19,nsw-is14,hnsw-tpami18}.
Among these, graph-based algorithms, such as HNSW~\cite{hnsw-tpami18} and NSG~\cite{nsg-vldb19}, have shown promising search performance over other approaches~\cite{survey-vldb21,survey-tkde19}. They efficiently explore neighborhood structures by constructing a graph where nodes represent feature vectors and edges denote potential nearest neighbor relationships.
Graph-based methods scale effectively with large datasets and adapt well to various data distributions.
As a result, they form the foundation of various ANNS services and vector databases, including 
ElasticSearch~\cite{Elasticsearch}, FAISS~\cite{faiss-tbd19}, Milvus~\cite{Milvus}, VSAG~\cite{vsag-vldb25},
and PostgreSQL~\cite{Postgresql-sigmod20}.

\begin{figure}
	\centering
	\includegraphics[width=\linewidth]{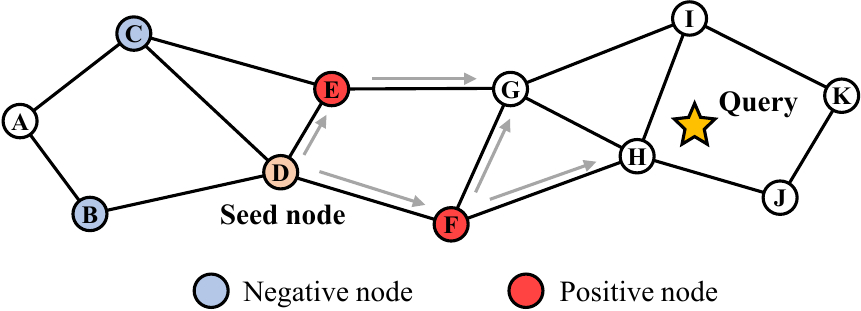}
	\caption{Illustration of the greedy search algorithm.
	}
	\label{fig:anns}
\end{figure}

The graph-based ANNS algorithms commonly use the greedy search algorithm for searching nearest neighbors. As shown in Figure~\ref{fig:anns}, it maintains a candidate set where candidates are assessed based on their distance from the query point. Starting from the specified seed node, each candidate's neighbors are examined. Neighbors whose distances to the query exceed the farthest distance in the candidate set are discarded (i.e., negative nodes), while others are included in the candidate set and further refined (i.e., positive nodes). The iterative distance calculation calls find nodes closer to the query, progressively moving towards the query point.

However, repeated distance calculations in high-dimensional spaces are costly and constitute the primary time bottleneck in ANNS. For example, our experiments show that distance calculations account for at least 83\% of the total running time across five widely used datasets with varying sizes and dimensionalities (\cref{background:running time analysis}). Furthermore, the majority of nodes involved in these calculations are negative objects, meaning they will not be included in the candidate set and do not affect the final result. Based on these observations, it suggests that developing a method to prune negative nodes could reduce expensive distance calculations without compromising accuracy.

Various routing techniques are proposed to improve the performance of graph-based ANNS.
The primary objective is to minimize distance computations during neighbor exploration.
For example, HCNNG~\cite{hcnng-pr19} and TOGG~\cite{togg-kbs21} employ KD-trees~\cite{kdtree-1975} to select points in the same direction as the query thereby only exploring a subset of neighbors for each search iteration. However, they tend to yield suboptimal query accuracy because of incorrect pruning, limiting their effectiveness.
FINGER~\cite{finger-www23} pre-computes and stores the residual vectors for all nodes' neighbors during construction, allowing for rapid distance estimation and node pruning during search.
Other ML-based optimizations~\cite{learning2route1-icml19,learning2route2-icml20,optlearning-sigmod20} learn routing functions, utilizing additional representations to facilitate optimal routing from the starting node to the nearest neighbor.
However, all these optimizations either compromise search accuracy or necessitate additional training and extra information, leading to increased construction time and memory footprint.

In this paper, we propose \pname{}, a novel routing strategy to guide
the navigation for graph-based ANNS. 
Compared to existing routing strategies, \pname{} achieves a favorable balance between search and construction efficiency.
\pname{} take inspiration from the characteristics of high-dimensional vector distribution:
\textit{in high-dimensional spaces, two random vectors are almost always very close to orthogonal}~\cite{geometry-1997}.
Following this theorem, in the triangles formed by the current search node, the query node, and the neighboring node, we observe that the angles associated with the current search node tend to concentrate around a specific value.
Based on this observation, \pname{} uses the cosine theorem to efficiently estimate the distance between the neighbor node and the query node.
The estimated distance is used to decide whether to prune the neighbor, thereby reducing unnecessary computational overhead. 
Since the estimated distance may have approximation errors, some nodes might be incorrectly pruned. 
\pname{} further introduces a technique to identify these incorrectly pruned nodes and recalculate their exact distances to the query to ensure high accuracy.
In summary, this paper makes the following contributions:
\begin{itemize}[left=0pt]
\item We conduct a thorough analysis of existing graph-based ANNS algorithms and find that all of them are plagued by repeated distance computations, which account for the majority of the overall search operation costs.
\item We propose \pname{}, a novel routing strategy that guides navigation by leveraging the characteristics of high-dimensional vector distributions. Collaborated with an error-correction technique, \pname{} effectively prunes massive unnecessary distance computations under the same accuracy criteria. 
\item We implement \pname{} and develop it as a plugin to enhance HNSW and NSG, two predominant graph-based ANNS algorithms. 
Our evaluation results show that \pname{} reduces the number of distance computations by up to 41.5\% while maintaining the same accuracy, thereby improving queries per second (QPS) by up to 1.48$\times$.
\end{itemize}


\section{Background and Motivation}
\label{background and motivation}

\subsection{Greedy Search}
The greedy search algorithm is commonly used in most graph-based methods for finding nearest neighbors. As shown in Algorithm~\ref{alg:search}, given a query point $q$ and a starting point $p$, the algorithm is designed to search for $efs$ nearest neighbors to $q$.
It maintains two priority queues: candidate queue $C$ that stores potential candidates to expand and top results queue $T$ that stores the current most similar candidates.
At each search iteration, it first extracts the current nearest point $c$ in $C$ and gets the furthest distance to the query $q$ from $T$ as an $upper\ bound$ (lines 3-4).
And then it visits $c$’s neighbors and computes their distance to $q$ respectively to expand the candidates.
For each neighbor, it checks whether its distance from $q$ is less than the $upper\ bound$ and if so, it pushes the neighbor into both $C$ and $T$, and updates the $upper\ bound$ simultaneously (lines 11-14).
The repeated distance calls constitute the primary time cost (line 11).

\begin{algorithm}[t]
    \caption{Greedy Search}
    \label{alg:search}
    \begin{algorithmic}[1]
		\Statex{\textbf{Input}: query $q$, starting point $p$, candidate neighbor limit $efs$}
		\Statex{\textbf{Output}: top results queue $T$}
		\State{candidate queue $C = \{ p \}, T = \{ p \}$, mark $p$ as visited}
		\While{$C$ is not empty}
			\State{$c$ ← nearest element from $C$ to $q$}
			\State{$upper\ bound$ ← farthest element distance from $T$ to $q$}
			\If{$dist(c, q) > upper\ bound$}
        		\State{return $T$}
       		\EndIf
			\For{each point $n$ $\in$ neighbors of $c$}
				\If{$n$ is visited}
					\State{continue}
				\EndIf
				\State{mark $n$ as visited}
				\If{$dist(n,q) < upper\ bound$ or $|T| < efs$}
					\State{$C.add(n), T.add(n)$}
					\If{$|T| > efs$}
						\State{$T.resize(efs)$, update $upper\ bound$}
					\EndIf
				\EndIf
			\EndFor
		\EndWhile
		\State{return $T$}
    \end{algorithmic}
\end{algorithm}

\begin{figure}[t]
	\centering
	\includegraphics[width=\linewidth]{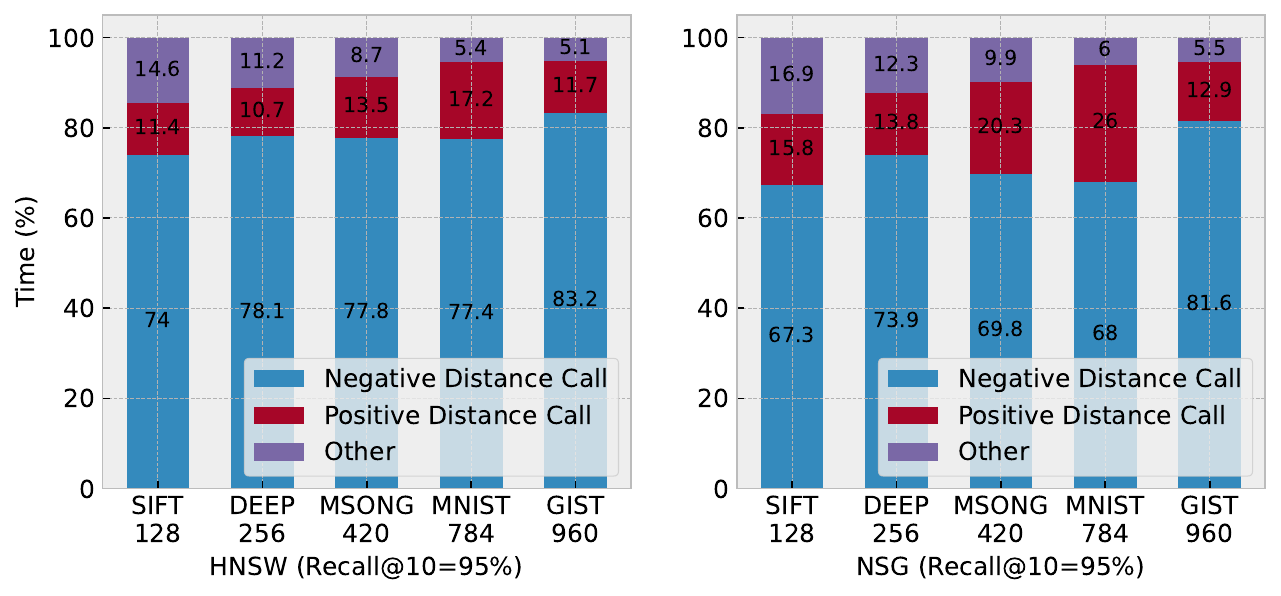}
	\caption{Running time analysis of the greedy search across five datasets with dimensions ranging from 128 to 960.
	}
	\label{bg:breakdown}
\end{figure}

\subsection{Running Time Analysis}
\label{background:running time analysis}
We measure the time consumption of greedy search across five public datasets using two state-of-the-art graph-based ANNS algorithms:
HNSW and NSG. 
In this test, if a node is added to the candidate queue after triggering a distance calculation, 
we refer to it as a positive node; otherwise, it is termed a negative node.

As shown in Figure~\ref{bg:breakdown},
when the data dimensionality increases from 128 to 960, 
the proportion of time spent on distance calculations rises from 85.4\% to 94.9\% on the HNSW algorithm, and from 83.1\% to 94.5\% on the NSG algorithm, respectively.
Moreover, we observe that the majority of nodes are negative, 
indicating that these nodes will not be added to the candidate set 
and will not have any further impact on the final result.
This suggests that if we can develop a method to prune these negative nodes,
we could significantly reduce expensive distance calculations without compromising accuracy.

\begin{figure}[t]
	\centering
	\includegraphics[width=2in]{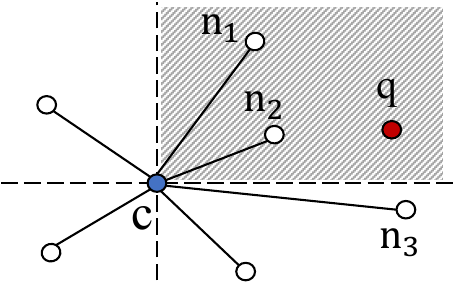}
	\caption{Optimized guided search in TOGG. Only the neighbors in the same direction as the query node $q$ will be considered, such as $n_1$ and $n_2$. However, this can easily overlook potentially valuable nodes, such as $n_3$.
	}
	\label{bg:togg}
\end{figure}

\begin{figure}[t]
	\centering
	\includegraphics[width=2in,height=1in]{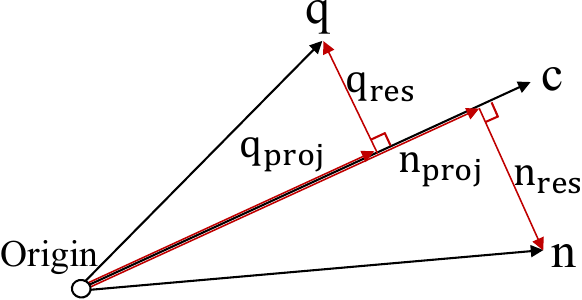}
	\caption{Decomposition of the distance computation in FINGER. The query node $q$ and the neighbor node $n$ can be represented by vectors parallel and orthogonal to the node $c$.}
	\label{bg:finger}
\end{figure}

\begin{table}[t]
	\caption{Comparison between different routing strategies.}
	\resizebox{\linewidth}{!}{
	\begin{tabular}{c|ccc}
	\hline
	Strategy & Construction time & Memory footprint & Accuracy \\ \hline
	TOGG       & Fast \ding{51}  & Low \ding{51}  &  Low  \ding{55}    \\
	FINGER  & Slow  \ding{55}  & High \ding{55}    & High   \ding{51}           \\
	\pname{}           & Fast \ding{51}  & Low \ding{51}  & High  \ding{51}       \\ \hline
	\end{tabular}
	}
	\label{table:comparsion}
\end{table}

\subsection{Existing Routing Strategies}
\label{background:existing routing strategies}
Several routing techniques have been proposed to minimize distance computations during neighbor exploration.
Here, we describe two state-of-the-art routing strategies,
including TOGG and FINGER,
which we use in the experimental section for comparison with \pname{}.

TOGG introduces a two-stage routing strategy that combines optimized guided search with a greedy algorithm. 
The routing process is explicitly divided into two detailed stages: (S1) the routing stage that is farther from the query and (S2) the routing stage that is closer to the query.
In stage S1, the primary focus is on quickly identifying the neighborhood of the query to enable prompt routing toward it. 
To achieve this, TOGG utilizes KD-trees to select points aligned with the query direction, thereby allowing it to explore only a subset of neighbors during each search iteration, as illustrated in Figure~\ref{bg:togg}.
In contrast, stage S2 emphasizes the need to thoroughly explore vertices near the query in order to obtain sufficiently accurate search results. Consequently, this stage not only examines the neighbors of each vertex in the candidate set but also investigates the neighbors of those neighbors. By relaxing the expansion constraint compared to traditional greedy algorithms, TOGG effectively mitigates the risk of getting trapped in local optima.
However, the utilization of KD-trees for navigation may result in reduced accuracy, as numerous potential nodes are filtered out during stage S1 and cannot be reintegrated in stage S2.

FINGER estimates the distance between each neighbor and the query. Specifically, for each node, it generates projected vectors locally for both neighbors and the query to define a subspace, and then applies Locality Sensitive Hashing (LSH)~\cite{lsh-02} to approximate the residual angles.
As illustrated in Figure~\ref{bg:finger}, the distance is decomposed as follows:
\begin{equation} 
	\Vert q-n\Vert^2 = \Vert q_{proj}-n_{proj}\Vert^2 + \Vert q_{res}\Vert^2 + \Vert n_{res}\Vert^2 - 2q_{res}^Tn_{res}
\end{equation}
During graph construction, FINGER will calculate and store $\Vert c\Vert$ (1 byte),  $\Vert n_{res}\Vert$ (4 bytes), projection coefficient $b$ 
($b = \frac{\Vert n_{proj}\Vert}{\Vert c\Vert}$, 4 bytes),
and other relevant LSH information for estimating the angle between $q_{res}$ and $n_{res}$.
Then, FINGER can quickly estimate distances and eliminate unnecessary calculations during query processing. 
However, this method necessitates additional computation and supplementary information, leading to increased time and memory overhead for graph construction.



\begin{figure}[t]
	\centering
	\includegraphics[height=1in]{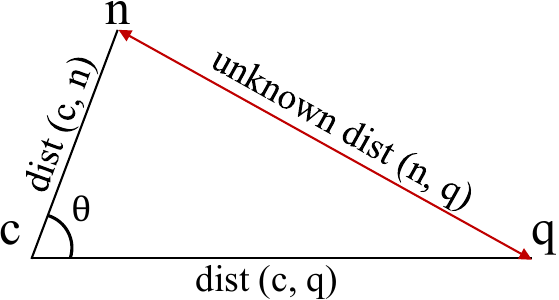}
	\caption{Triangle formed by the current search node $c$, a query $q$, and a specific neighbor of $c$, referred to as $n$. Our object is to estimate $dist(n,q)$ with low cost.
	}
	\label{bg: triangle}
\end{figure}

\begin{figure}[]
	\centering
	\includegraphics[width=\linewidth]{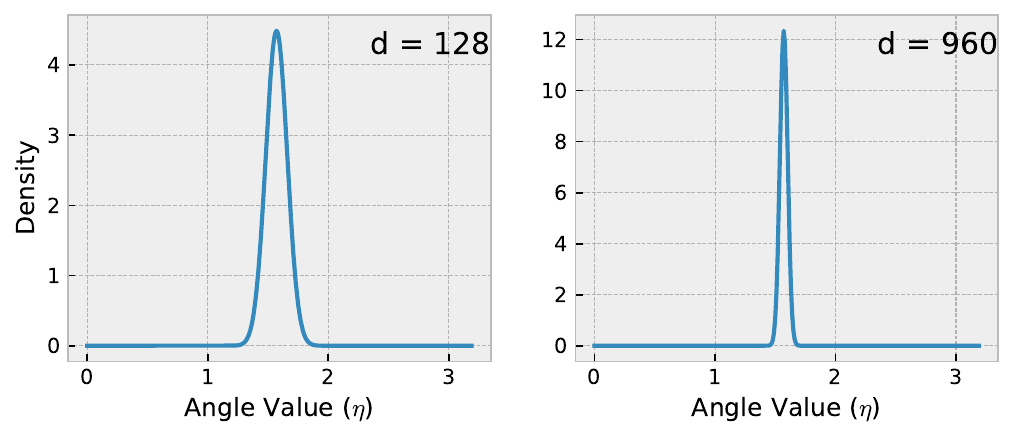}
	\caption{Probability density of angles between two random vectors, where $d$ represents the vector dimension.
	}
	\label{bg:probability_density1}
\end{figure}

Existing routing strategies either compromise search accuracy or require additional construction overhead.
In contrast, \pname{} strikes a favorable balance between search and construction efficiency.
Table~\ref{table:comparsion} compares \pname{} with existing routing strategies, and the detailed experimental results are presented in~\cref{eva:comparison_to_other_strategies}.

\section{Problem Definition and Analysis}
\label{inspiration}
In this section, we present the insight behind \pname{}, 
an effective distance estimation method that leverages the characteristics of high-dimensional vector distributions.

\begin{figure*}[t]
	\centering
	\includegraphics[width=\linewidth]{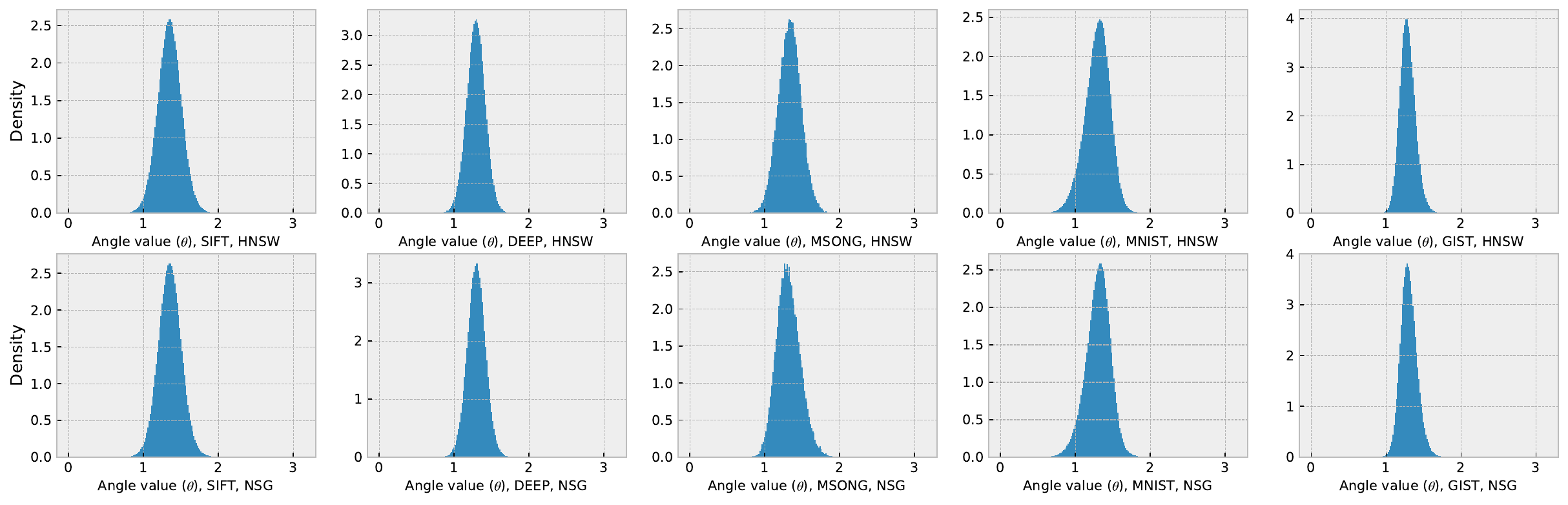}
	\caption{Probability density of angles on HSNW and NSG algorithms. The number of queries is 0.1\% of the dataset size.}
	\label{angle_distribution}
\end{figure*}

\begin{figure*}[t]
	\centering
	\includegraphics[width=\linewidth]{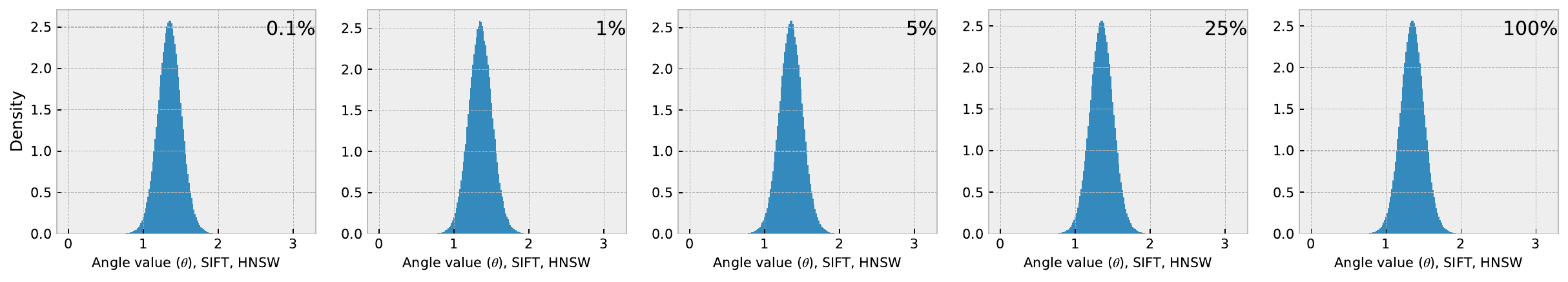}
	\caption{Probability density of angles on HNSW algorithm. The number of queries varies from 0.1\% to 100\% of the dataset size.}
	\label{angle_distribution_with_different_query_number}
\end{figure*}

\subsection{Problem Statement}
As shown in Figure~\ref{bg: triangle}, during each search iteration, given a query $q$, a current search node $c$, and a specific neighbor of $c$ called $n$, traditional greedy search algorithms compute the exact Euclidean distance between $q$ and $n$ (i.e., $dist(n,q)$), with a time complexity of $O(d)$, where $d$ is the data dimension. 
Our objective is to approximate this distance. 
If the estimated distance exceeds the $upper\ bound$ (i.e., the furthest distance to the query $q$ from the top results queue $T$ as stated in Algorithm~\ref{alg:search}), then node $n$ can be pruned. 
This approach ensures that only nodes with a potential to be closer to the query $q$ are considered, thus optimizing the search process and reducing computational overhead.

\subsection{Naive Solution: Triangle Inequality}
Currently, we only know $dist(c, q)$ because $c$ is extracted from the candidate queue, and $dist(c, q)$ is pre-calculated in a previous search iteration. 
However, this information alone is insufficient to estimate $dist(n, q)$. 
Note that $n$ is a neighbor of $c$, so $dist(c, n)$ was calculated during the graph construction and can be retained in memory.

After obtaining the exact lengths of the edges (i.e., $dist(c,q)$ and $dist(c,n)$), we can directly apply the triangle inequality to estimate the lower bound of the third edge's length. 
Specifically, in $\triangle cnq$:

\begin{equation} 
dist(n,q) \geq |dist(c,n) - dist(c,q)|
\end{equation}

And if $|dist(c,n) - dist(c,q)|$ is greater than the $upper\ bound$, 
we can directly prune node $n$.
However, our research shows that this method is not always effective. 
For instance, it reduces the number of distance computations by only 0.08\% using the SIFT dataset on the HNSW algorithm, which is negligible in the overall search process.
Although the triangle inequality provides an accurate lower bound, it is often too loose to effectively filter out nodes.

\subsection{Optimized Solution: Cosine Theorem}
\label{bg:opt_solution}

To obtain a more accurate distance estimation, additional information, the angle $\angle ncq$ ($\theta$ in Figure~\ref{bg: triangle}), is required.
With this angle, we can then calculate $dist(n, q)$ using the cosine theorem.

The distribution of vectors in high-dimensional spaces has a unique property:
\textit{in high-dimensional spaces, two random vectors are almost always
very close to orthogonal.}
The probability density function of the angle $\eta$ between two randomly selected vectors can be formulated as follows:

\begin{equation} 
\label{equation1}
P(\eta)=\frac{\Gamma\left(\frac{d}{2}\right)}{\Gamma\left(\frac{d-1}{2}\right) \sqrt{\pi}} \sin ^{d-2} \eta
\end{equation}
where $d$ is the dimensionality of the vector and $\Gamma()$ is the Gamma function. 
This indicates that the angle distribution is solely determined by the dimensionality of the vectors.
To better understand, Figure~\ref{bg:probability_density1} plots the probability density for $d = 128$ and $d = 960$. As $d$ increases, the values of $\eta$ become increasingly concentrated around $0.5\pi$ (orthogonal).

Inspired by this, we question whether the angle $\theta$ between vectors $\overrightarrow{cn}$ and $\overrightarrow{cq}$ in Figure~\ref{bg: triangle} satisfies the above equation.
Because the position of neighbor $n$ is completely random with respect to the current node $c$,
the direction of vector $\overrightarrow{cn}$ relative to $\overrightarrow{cq}$ is independent.
Therefore, the values of angle $\theta$ may also exhibit special distribution characteristics.  

To further verify this, we record the values of $\theta$ along the search paths across different datasets and algorithms using randomly generated queries.
The number of queries varies from 0.1\% to 100\% of the number of nodes to construct the graph.
From the results in Figure~\ref{angle_distribution} and Figure~\ref{angle_distribution_with_different_query_number}, we have the following observations.
(1) We observe that the values of $\theta$ follow a skewed distribution, with the center around $0.5\pi$, which is similar to the trend shown in Figure~\ref{bg:probability_density1}.
(2) We observe that the distribution of angles is determined solely by the dataset and remains unaffected by the graph construction algorithm or the number of query nodes.
 This indicates that the angle distribution is an intrinsic property of the dataset.
Because the dataset's vector dimensionality and distribution remain constant, the angle distribution can be used indefinitely once computed.

Based on these observations, we propose the key design concept of \pname{}.
For each dataset, \pname{} selects a specific angle derived from its angle distribution (i.g., the central point) to set the value of $\theta$ in the cosine theorem. 
Given that the angle exhibits a skewed characteristic, using a single representative value for all the angle values along the search path enables a efficient estimate of $dist(n,q)$ while keeping the distance estimation error within a small range.

\section{\pname{} Design}
\label{sec:design}
This section describes the graph construction and query processes of \pname{}. \pname{} is designed as a plugin to improve the query performance of existing graph-based ANNS algorithms.

\subsection{Graph Construction}
\textbf{Acquisition of additional information.} 
As a prerequisite for \pname{}, we need to obtain two additional pieces of information during graph construction: (1) the distances between each node and all its neighbors, and (2) the angle distribution of $\theta$.
The distances between each node and its neighbors are already 
calculated during the graph construction process in all
the existing graph-based ANNS algorithms.
\pname{} simply saves them in memory rather than discarding them.
To obtain the angle distribution, we will randomly generate $n_{sample}$ query nodes after the graph has been constructed.
On the search paths of these nodes, we simultaneously use the cosine theorem to calculate the angle values.

\textbf{Extra time overhead.}
By default, the value of $n_{sample}$ is set to 0.1\% of the number of nodes used to construct the graph.
Because the sampling frequency is sufficient to obtain an accurate angle distribution (as shown in Figure~\ref{angle_distribution_with_different_query_number}), while the computational overhead remains negligible.
Our experiments show that the additional construction time does not exceed 4\% (\cref{eva:comparison_to_other_strategies}).

\textbf{Extra memory overhead.}
One concern is that storing distances between neighbors 
incurs extra memory overhead. In graph-based ANNS algorithms,
memory is used for storing high-dimensional vectors, graph indexs,
and distance values ($mem_{dist}$) saved by \pname{}.
Our study shows that $mem_{dist}$ leads to an additional 2\% to 21\%
overhead compared to the approaches without using \pname{} (\cref{eva:comparison_to_other_strategies}).
The ratio of additional memory consumption is decreased as the dimensionality of datasets is increased.

\subsection{Graph Search}
\label{design: search}
\textbf{Pruning strategy.}
By utilizing the stored $dist(c,n)$ and $dist(c,q)$, 
along with the angle $\theta$ obtained from the distribution of high-dimensional random vectors (\cref{bg:opt_solution}),
we can efficiently estimate $dist(n,q)$ using the cosine theorem.
If $dist(n,q)$ is not less than the $upper\ bound$,
we can avoid making the expensive distance call to calculate the exact distance.
Otherwise, the normal routing process will proceed as usual.
Adjusting the angle $\theta$ allows us to control the threshold for the pruning strategy. 
Specifically, larger angle values lead to more nodes being pruned, but they also increase the estimation error, impacting accuracy. 
Thus, there is a trade-off between accuracy and the number of distance calculations. 
We evaluate it in detail in~\cref{eva:sensitive analysis}.

\begin{figure}[t]
	\centering
	\includegraphics[height=1in]{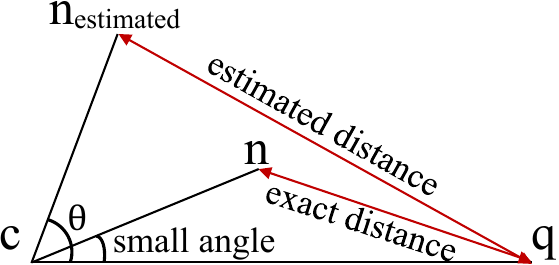}
	\caption{Example of incorrect estimation.
	}
	\label{fig:corrective}
	\vspace{-0.1in}
\end{figure}

\textbf{Error correction.} 
Although many unnecessary nodes can be effectively pruned, a significant issue remains. 
Our pruning strategy may introduce approximation errors, which may lead to the 
removal of important nodes and notably impact the final results.
For instance, as illustrated in Figure~\ref{fig:corrective}, given a query $q$ and the current search node $c$,
we expand the search by exploring the neighbors of $c$. 
The specific neighbor node $n$, which lies in the direction of $q$ (i.e., with an extremely small angle), is closer to $q$ than $c$, making it a strong candidate for refining the final results.
However, for such neighbors, the error in our estimated distance compared to the actual distance can be significant, raising the likelihood that these promising nodes are mistakenly pruned. Additionally, the convergence paths from these neighbors may also be overlooked, negatively impacting accuracy.

To address this issue, we devise an error correction mechanism to identify 
erroneously deleted nodes. 
The key idea is that nodes mistakenly pruned and located near the query node are likely to be revisited in subsequent searches because of the graph's connectivity.
This means that even if some positive nodes are pruned initially, they can still be reached through alternative paths in later stages of the search.
For example, as shown in Figure~\ref{fig:anns}, if node $G$ is pruned while evaluating the neighbors of node $E$, it will still be revisited when expanding from node $F$.
Based on this idea, if a node has been pruned by \pname{}, any 
future access to that node should involve calculating its exact distance to ensure high accuracy.


\textbf{Put it all together.}
Algorithm~\ref{alg:search_ours} describes the graph search process of \pname{}.
Compared to the existing ANNS algorithms using greedy search (e.g., 
Algorithm~\ref{alg:search}), \pname{} uses the approximated
distance to aggressively prune nodes (lines 10-13)
before each exact distance calculation (line 15).
Errors may happen when a node is marked as ``pruned'' (line 12).
In this case, when the node is
revisited in the future search, \pname{} will mark the node $n$ 
as ``visited'' and calculate the exact $dist(n,q)$ (lines 14-15).
We design \pname{} as a plug-in that can be applied on top of 
any existing graph-based ANNS algorithms that utilize greedy search,
requiring only minimal changes to the existing code.

\begin{algorithm}[tb]
    \caption{Graph Search of \pname{}}
    \label{alg:search_ours}
    \begin{algorithmic}[1]
		\Statex{\textbf{Input}: query $q$, starting point $p$, candidate neighbor limit $efs$}
		\Statex{\textbf{Output}: top results queue $T$}
		\State{candidate queue $C = \{ p \}, T = \{ p \}$, mark $p$ as visited}
		\While{$C$ is not empty}
			\State{$c$ ← nearest element from $C$ to $q$}
			\State{$upper\ bound$ ← farthest element distance from $T$ to $q$}
			\If{$dist(c, q) > upper\ bound$}
        		\State{return $T$}
       		\EndIf
			\For{each point $n$ $\in$ neighbors of $c$}
				\If{$n$ is visited}
					\State{continue}
				\EndIf

				\If{$n$ is not pruned and $|T| \geq efs$}
				\If{$appx\_dist(n, q) \geq upper\ bound$}
					\State{mark $n$ as pruned}
					\State{continue}
				\EndIf
				\EndIf
				
				\State{mark $n$ as visited}
				\If{$dist(n,q) < upper\ bound$ or $|T| < efs$}
					\State{$C.add(n), T.add(n)$}
					\If{$|T| > efs$}
						\State{$T.resize(efs)$, update $upper\ bound$}
					\EndIf
				\EndIf
			\EndFor
		\EndWhile
		\State{return $T$}
    \end{algorithmic}
\end{algorithm}

\subsection{Applicability to Other Distance Metrics}
\label{sec:discussion}
In addition to Euclidean distance, \pname{} can be easily extended to two other commonly used distance metrics: inner-product and cosine distance, through simple transformations. 
For a query $q$ and data point $c$ with an angle $\theta$ between them,
the inner-product distance is given by: $IPDist(c,q) = 1 -\Vert c\Vert \Vert q\Vert cos \theta$. 
The relationship between Euclidean and inner product distance is:
\begin{equation}
    \begin{split}
    EuclideanDist(c,q)^2 &= \|c\|^2+\|q\|^2-2\|c\|\|q\|cos \theta \\
    &= \|c\|^2+\|q\|^2+2\ IPDist(c,q)-2
    \end{split}
\end{equation}
Since the calculation of $\|q\|$ is a one-time operation for each query, it remains computationally inexpensive. 
Moreover, $\|c\|$ (i.e., the norm for each inserted vector) can be precomputed and stored during the graph construction phase. 
The associated space and time costs are acceptable.
Specifically, the space required for storing the norm of a vector (typically with dimensions greater than 100) 
is less than 1\%. The computational cost of calculating the norm one time is negligible compared to the numerous distance calculations required for graph construction.
In addition, the cosine distance is equivalent to the inner-product distance on normalized data and query vectors~\cite{adsampling-sigmod23}, where \pname{} is still applicable. 
We evaluate the generality of \pname{} in~\cref{eva:scalability and generality}.
\begin{table}[t]
    \caption{Data statistics.}
    \begin{tabular}{c|cccc}
    \hline
    Dataset & \# Base        & \# Query            & \# Dim                  & Type                 \\ \hline
    SIFT~\cite{sift-gist}    & 1,000,000    &  10,000      & 128                  & Image                \\
    DEEP~\cite{deep}    & 1,000,000    &  1,000      & 256                  & Image                \\
    MSONG~\cite{msong}   & 992,272   & 200          & 420 & Audio \\
    MNIST~\cite{mnist}   & 60,000     &   10,000       & 784 &   Image                   \\
    GIST~\cite{sift-gist}    & 1,000,000  &   1,000       & 960 &    Image                  \\ \hline
    \end{tabular}
    \label{datasets}
\end{table}

\section{Evaluation}
\label{sec:evalution}

Our experiments involve three folds.
First, we compare our methods with traditional graph-based approaches to empirically demonstrate that \pname{} can be integrated into graph-based greedy search methods for improved performance. 
Second, we evaluate the sensitivity, generality, and scalability of \pname{} in various scenarios.
Third, we compare our methods with other routing strategies to assess their efficiency and effectiveness.

\subsection{Experimental Setting}
\label{eva:experimental setting}
\textbf{Implementation setup.}
We run the experiments on a Linux
server with two Intel Xeon Gold
5318Y CPUs. Each CPU has 24 physical/48 logical cores,
\SI{64}{GB} DRAM.
All source codes are compiled with g++10.3 with -O3 optimization.
Following previous work~\cite{adsampling-sigmod23,survey-vldb21,survey-tkde19}, we disable all hardware-specific optimizations 
such as SIMD and memory prefetching so as to focus on the comparison among algorithms themselves.
To improve construction efficiency, the code involving vector calculation is parallelized for the index construction.
All tests are evaluated on a single thread by default.

\textbf{Datasets and metrics.}
We use five public datasets with varying sizes and dimensionalities,
detailed in Table~\ref{datasets}.
They are commonly employed to benchmark ANNS algorithms~\cite{annbenchmarks-is20}.
We use the Euclidean distance as the distance metric by default.
The comparison metrics used to evaluate performance are:
(1) Recall@K: it measures search accuracy by comparing the approximate point set $T$ found for a given query $q$ with the true $K$ nearest neighbor result set $R$. 
It is defined as: $Recall@K = |T \cap R|/K$. 
By default, $K$ is set to 10.
(2) QPS: it indicates the number of queries a machine can handle per second.
(3) Speedup in distance calls~\cite{togg-kbs21,finger-www23,nhq-nips24}: it measures the efficiency of a method by comparing the number of distance calculations. It is defined as the ratio of distance calculations in the brute-force method to those in the optimized routing method.
Since the accuracy and performance of greedy search can vary on the size of the candidate array $efs$ (see Algorithm~\ref{alg:search}), we adjust $efs$ to plot both the recall–QPS curve and the recall–speedup curve.

\textbf{Target comparisons.}
We implement \pname{} on HNSW\footnote{\url{https://github.com/nmslib/hnswlib}} and NSG\footnote{\url{https://github.com/ZJULearning/nsg}}, two widely used graph indexes for ANNS in production services like Facebook AI Research~\cite{faiss-tbd19} and Alibaba e-commerce search~\cite{nsg-vldb19}.
We follow the default parameter settings of their public code.
For HNSW, the neighbor limit $M$ for each vector is set to 32 and the insertion candidate neighbor limit $efc$ is set to 256.
For NSG, the neighbor limit $R$ is set to 70, the insertion candidate neighbor limit $C$ is set to 500, and the insertion priority queue length $L$ is set to 60.
We first compare \pname{} with initial HNSW and NSG algorithms to demonstrate how \pname{} accelerates them.
Then, we compare \pname{} with two currently leading-edge routing strategies including
(1) TOGG\footnote{\url{https://github.com/whenever5225/TOGG}}, which uses KD-trees to select querying neighbor nodes, and add a fine-tuned step when searching nodes near the query and
(2) FINGER\footnote{\url{https://github.com/Patrick-H-Chen/FINGER/tree/main}}, which approximates the distance function in graph-based methods by estimating angles between neighboring residual vectors.
Moreover, we evaluate two variants of our systems to distinguish the effect of each proposed design component. Specifically, \pname{}\_O refers to the version that only enables the pruning strategy while \pname{} incorporates both pruning and error-correction strategies.

\begin{figure*}[]
    \centering
    \includegraphics[width=\linewidth]{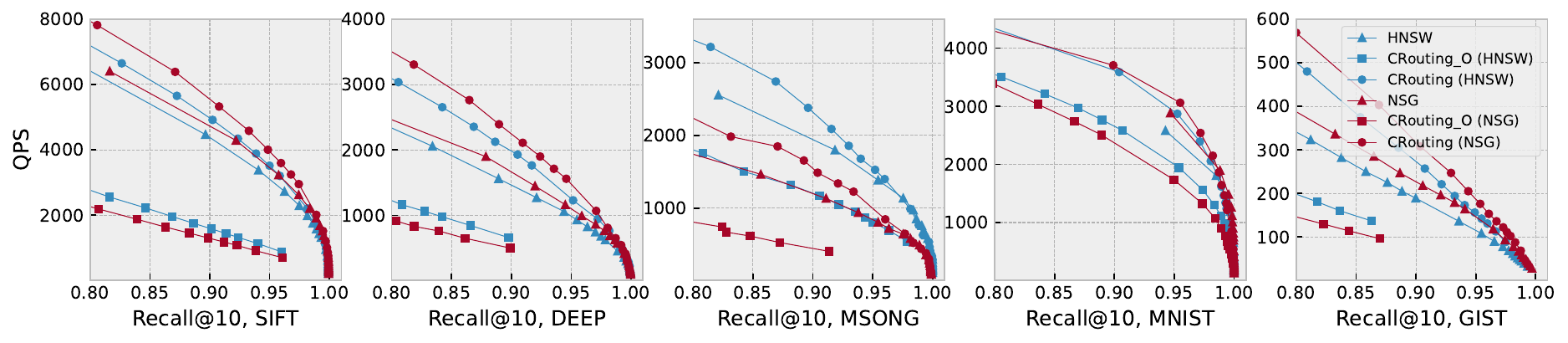}
    \caption{Recall-QPS curves (the top right is better).}
    \label{fig:recall-qps-anns}
\end{figure*}

\begin{figure*}
	\centering
	\includegraphics[width=\linewidth]{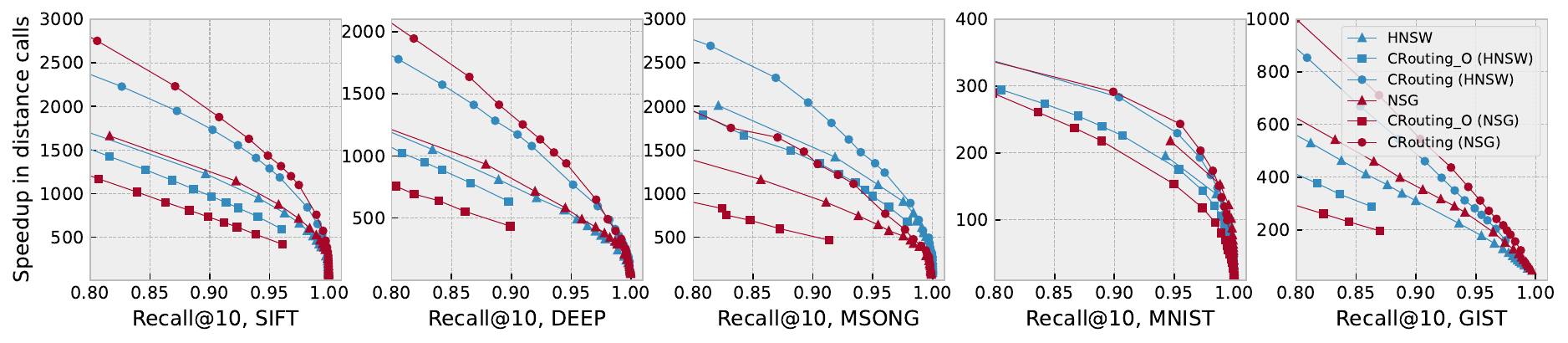}
	\caption{Recall-speedup curves (the top right is better).}
	\label{fig:recall-speedup-anns}
\end{figure*}

\begin{figure}
	\centering
	\includegraphics[width=\linewidth]{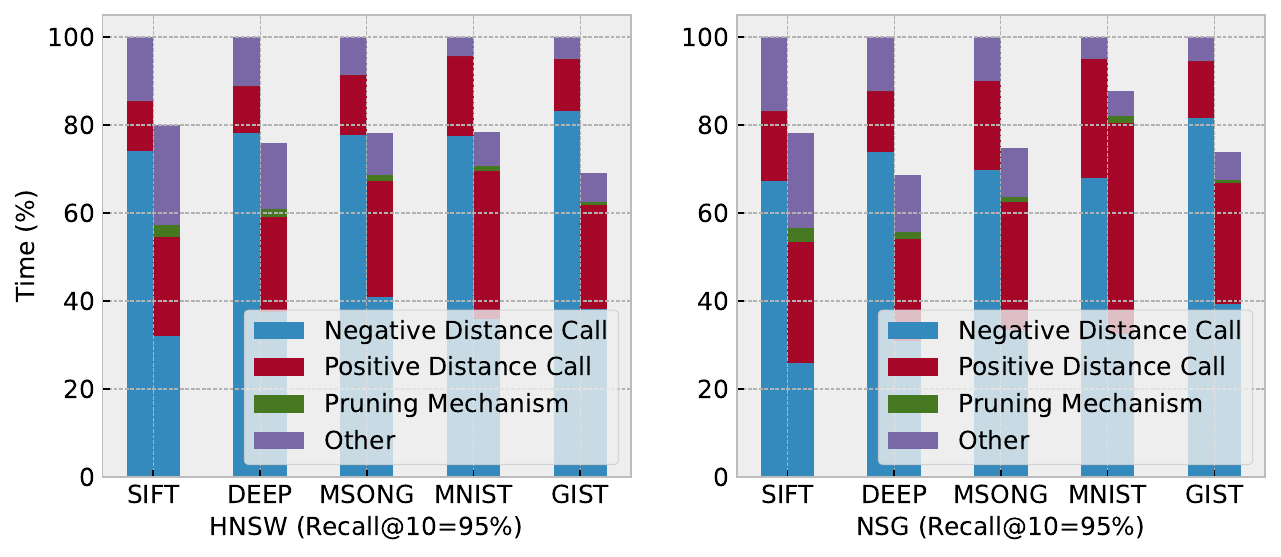}
	\caption{Time cost breakdown. On each dataset, the bar on the left represents the original ANNS algorithms, while the bar on the right represents \pname{}. The time cost is normalized by the cost of the original algorithms.
	}
	\label{fig:evaluation_time_breakdown}
\end{figure}

\subsection{Improvements over ANNS Algorithms}
Figure~\ref{fig:recall-qps-anns} plots the recall-QPS curves on five datasets.
We focus only on the region with the recall at least 80\% based on practical needs.
Overall, we can observe clearly that \pname{} exhibits the best performance across all datasets. 
\pname{} improves the QPS of HNSW and NSG by 1.12$\times$ to 1.48$\times$ and 1.11$\times$ to 1.47$\times$, respectively.
This demonstrates the effectiveness of our methods.
We also find that the performance of \pname{}\_O is significantly poor, even falling below the original algorithms. 
This indicates that relying solely on the pruning strategy is insufficient, as it results in the incorrect deletion of a large number of nodes.

To further analyze the reasons behind the performance improvement,
we show the detailed recall-speedup curves in Figure~\ref{fig:recall-speedup-anns}.
The recall-speedup curves display a trend similar to the recall-QPS curves.
\pname{} achieves a speedup  in distance calls ranging from 1.22$\times$ to 1.58$\times$ compared to HNSW, and from 1.15$\times$ to 1.71$\times$ compared to NSG, indicating a significant reduction in the number of distance calculations, thereby enhancing the QPS.

Moreover, we decompose the time cost when targeting 95\% recall in Figure~\ref{fig:evaluation_time_breakdown}.
We observe that \pname{} reduces the distance calculation time for negative nodes by 47.4\% to 61.6\%, supporting our motivation discussed in~\cref{background:running time analysis}.
While \pname{} introduces additional checks for the pruning mechanism, it still reduces the overall running time by 12.4\% to 31.4\%.
The pruning mechanism accounts for only 0.7\% to 3.9\% of the total time.
This low overhead is primarily due to the fact that a single cosine calculation involves only a few multiplications and additions.
In contrast, calculating the exact distance for a $d$-dimensional vector
requires $d$ multiplications, $d-1$ additions, and a single random memory read operation.
Thus, the pruning algorithm results in significant performance gains.

\begin{table}[t]
	\small
	\centering
	\caption{Performance with varying candidate array sizes on the DEEP dataset. $Hop$ means the number of distance calls.}
	\label{table:ablation study}
	\begin{tabular}{@{}l*{3}{ll}@{}}
	\toprule
	\multirow{2}{*}{efs} & 
	\multicolumn{2}{c}{HNSW} & 
	\multicolumn{2}{c}{\pname{}\_O} & 
	\multicolumn{2}{c}{\pname{}} \\
	\cmidrule(lr){2-3} \cmidrule(lr){4-5} \cmidrule(lr){6-7}
	 & {Recall} & {Hop} & {Recall} & {Hop} & {Recall} & {Hop} \\
	\midrule
	30   &  \cellcolor{blue!20}{0.889}  &  \cellcolor{blue!20}{1234651}  & 0.204  & 233588  & 0.676  & 413111  \\
	40   & 0.921  & 1507433  & 0.256  & 253973  & 0.753  & 489692  \\
	60   & 0.954  & 2032356  & 0.340  & 291073  & 0.842  & 635833  \\
	80   & 0.970  & 2534864  & 0.407  & 325629  & \cellcolor{blue!20}{0.886}  & \cellcolor{blue!20}{779403}  \\
	\rowcolor{red!20} 
	100  & 0.978  & 3013367  & 0.453  & 358959  & 0.917  & 926499  \\
	200  & 0.994  & 5194785  & 0.620  & 518164  & 0.972  & 1678470 \\
	300  & 0.997  & 7143884  & 0.708  & 672941  & 0.986  & 2431958 \\
	400  & 0.998  & 8934851  & 0.772  & 825041  & 0.992  & 3174724 \\
	500  & 0.999  & 10624804 & 0.808  & 977380  & 0.995  & 3909369 \\
	700  & 0.999  & 13772090 & 0.866  & 1279024 & 0.998  & 5343785 \\
	900  & 0.999  & 16686444 & \cellcolor{blue!20}{0.897}  & \cellcolor{blue!20}{1578669} & 0.998  & 6736528 \\
	\bottomrule
	\end{tabular}
	\end{table}

\subsection{Ablation Study}
\label{eva:ablation study}
The accuracy and performance of the greedy search can vary depending on the size of the candidate array $efs$ (see Algorithm~\ref{alg:search}). Table~\ref{table:ablation study} illustrates the recall@10 and hop (i.e., the number of distance calls) for different values of $efs$. 
When $efs$ equals 100 (highlighted in pink), \pname{}\_O reduces the number of hops by 90\% of the original greedy search algorithm, but at the cost of a significant drop in recall (from 97\% to 45\%).
In contrast, enabling the error-correction technique allows \pname{} to achieve 91\% recall. 
Although this results in more hops than \pname{}\_O, \pname{} still reduces the number of hops by 70\% compared to the original greedy search.

The pruning strategy sacrifices accuracy for performance, ensuring a substantial reduction in the number of distance calls. On the other hand, the error-correction strategy sacrifices performance for improved accuracy. 
Because the revisited nodes have a higher likelihood of being positive nodes, it is worthwhile to recalculate their distances.
By combining both techniques, \pname{} achieves a better balance between performance and accuracy. For instance, when recall reaches around 89\% (highlighted in purple), \pname{} reduces the number of hops by 36.9\% compared to the original greedy search, and by 50.6\% compared to \pname{}\_O.

\subsection{Error Analysis}
\label{eva:error analysis}
Although we cannot mathematically prove the approximation error bounds of \pname{},
we assess two accuracy metrics: relative error and the incorrect pruning ratio, to empirically verify that its accuracy is acceptable. 
Table~\ref{tab:avg_error} presents the relative error between the estimated distance $D_{estimated}$ and the true distance $D_{true}$. 
The relative error is defined as $e = \frac{|D_{true} - D_{estimated}|}{D_{true}}$.
The average relative error of \pname{} is approximately 6\%.
Additionally, Table~\ref{tab:false_negative} displays the number of false negatives, where nodes incorrectly classified as negative are actually positive.
The maximum ratio of incorrect pruning remains below 6\% across all datasets.
These results demonstrates the effectiveness of \pname{} in distance estimation.

\begin{table}[t]
\small
\centering
\caption{Average relative error of \pname{}.}
\label{tab:avg_error}
\begin{tabular}{l|ccccc}
\hline
 & SIFT & DEEP & MSONG & MNIST & GIST \\
\hline
HNSW & 7.00\% & 6.25\% & 6.71\% & 6.82\% & 5.14\% \\
NSG  & 6.84\% & 5.54\% & 6.47\% & 6.93\% & 5.07\% \\
\hline
\end{tabular}
\end{table}

\begin{table}[t]
\small
\centering
\caption{The ratio of incorrect pruning of \pname{}.}
\label{tab:false_negative}
\begin{tabular}{l|ccccc}
\hline
 & SIFT & DEEP & MSONG & MNIST & GIST \\
\hline
HNSW & 3.91\% & 4.16\% & 3.46\% & 2.44\% & 5.83\% \\
NSG  & 3.46\% & 3.24\% & 4.70\% & 2.48\% & 5.37\% \\
\hline
\end{tabular}
\end{table}





\begin{figure}[bt!]
	\centering
	\subfigure[\pname{}\_O]{
		\begin{minipage}[b]{0.97\linewidth}
		\includegraphics[width=\linewidth]{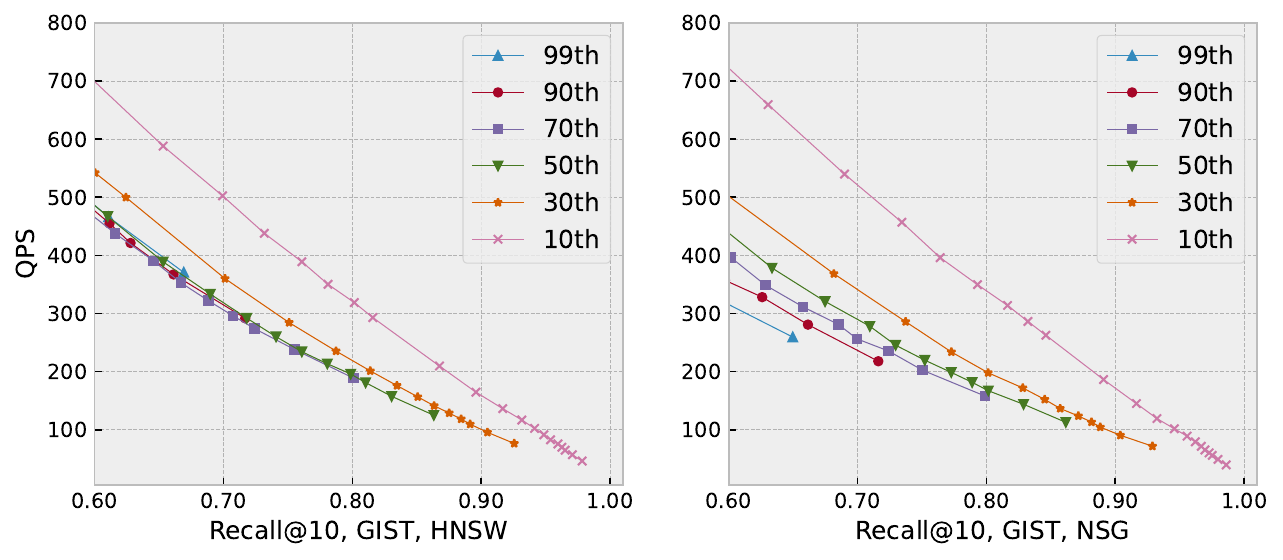}
		\end{minipage}
	}
	\subfigure[\pname{}]{
		\begin{minipage}[b]{0.97\linewidth}
		\includegraphics[width=\linewidth]{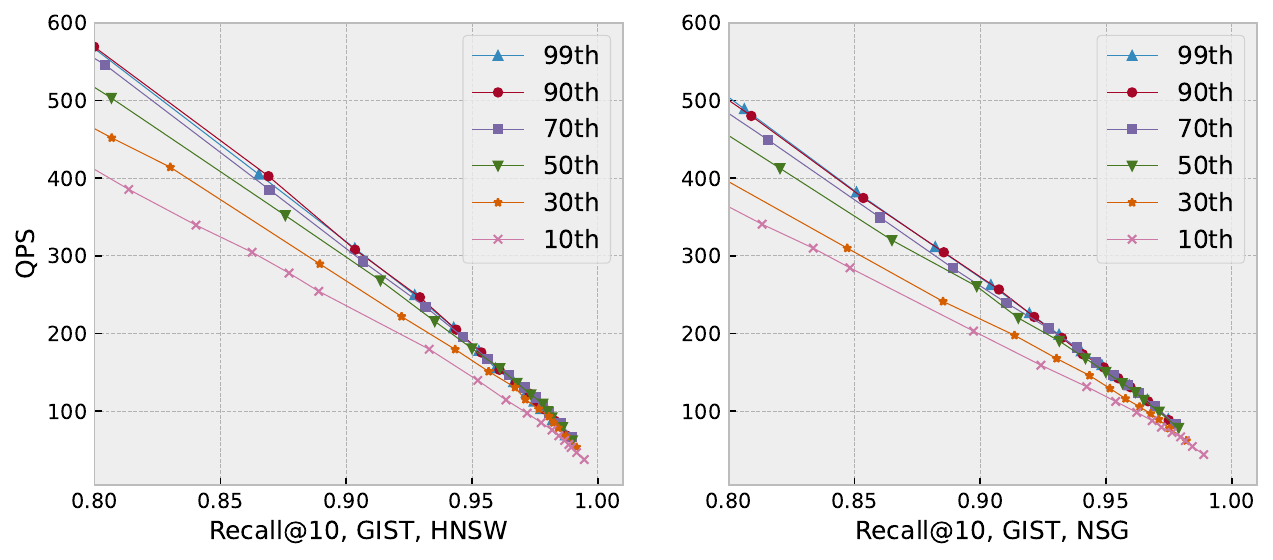}
		\end{minipage}
	}
	\caption{Effect of the pruning threshold.}
	\label{fig:effect_of_pruning_threshold}
\end{figure}

\begin{figure}[t]
	\centering
	\includegraphics[width=\linewidth]{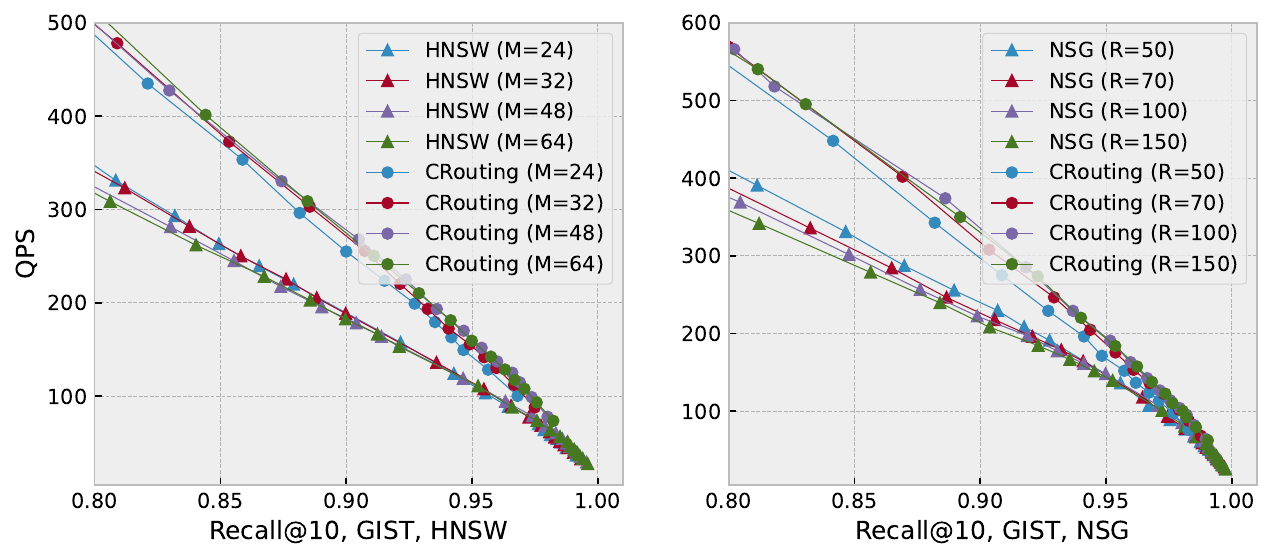}
	\caption{Effect of the number of connected neighbors. 
	}
	\label{fig:effect_of_neighbors}
\end{figure}

\begin{figure}
	\centering
	\includegraphics[width=\linewidth]{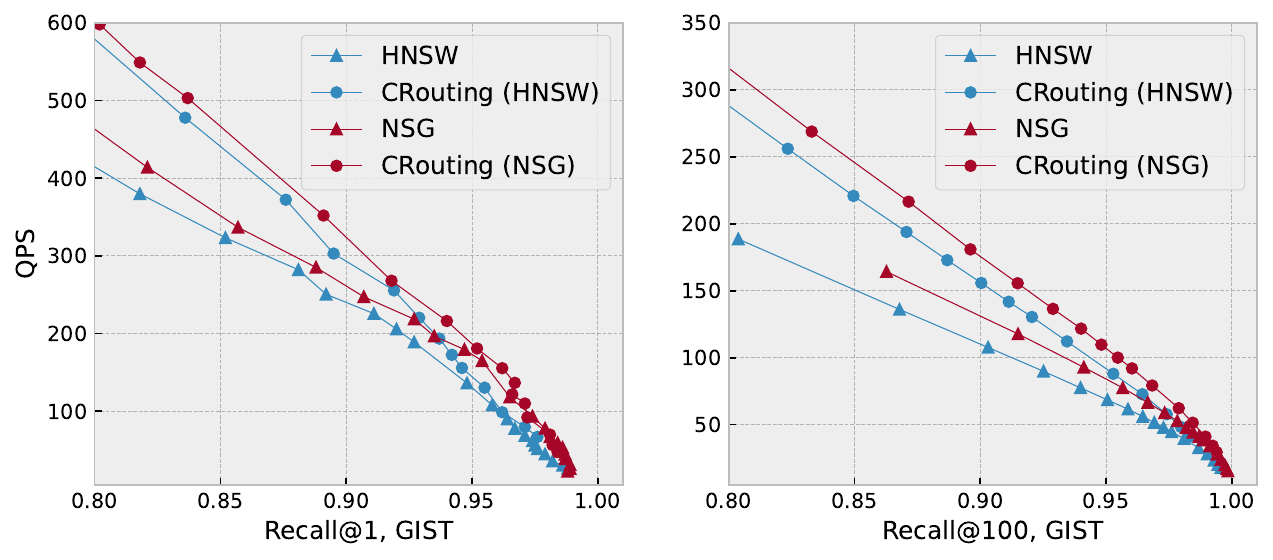}
	\caption{Effect of different result numbers. 
	}
	\label{fig:effect_of_K}
	\vspace{-0.1in}
\end{figure}

\subsection{Sensitive Analysis}
\label{eva:sensitive analysis}
The only parameter introduced by \pname{} is the pruning threshold (i.e., the value of $\theta$).
Other parameters, such as the number of neighbors,  $M$ and $R$, and result number $K$,
are inherent to the graph-based ANNS algorithms and are set based on dataset dimensions, accuracy requirements, and latency constraints.

\textbf{Effect of the pruning threshold.}
Figure~\ref{fig:effect_of_pruning_threshold} illustrates the impact of varying the pruning threshold by adjusting the angle from the 10th to the 99th percentile of the angle distribution. 
Larger angles for \pname{}\_O  generally result in lower performance. 
In contrast, for \pname{} with error-correction technique enabled,
selecting a larger angle typically leads to improved performance.
This is because larger angles aggressively prune more nodes, which enhances the QPS.
Although this increases the likelihood of mistakenly pruning some nodes, the accuracy loss from such errors can be mitigated by our error-correction mechanism.
Across all datasets and algorithms, we consistently observe
the best performance at the 90th percentile.
Consequently, we set the pruning threshold to the angle value at 90th percentile.

\textbf{Effect of the number of neighbors.}
The parameters $M$ and $R$ control the number of connected neighbors on HNSW and NSG algorithms, respectively, and this number directly influences the overall size of the graph.
In practical applications, users often adjust the number of neighbors based on their memory capacity and recall requirements.
Figure~\ref{fig:effect_of_neighbors} shows the recall-QPS curves of \pname{} on the GIST dataset with varying numbers of neighbors.
\pname{} consistently enhances QPS in all scenarios.
Additionally, we find that as the number of neighbors increases, the performance improvements brought by \pname{} become more pronounced.
As the dimensionality of real-world data increases, more complex and intricate graphs are needed to achieve high recall~\cite{hnsw-tpami18,nsg-vldb19}. 
This inevitably increases the number of connected neighbors, making our approach more attractive.

\textbf{Effect of the result number.} 
The parameter $K$ in Recall@K represents the result number. 
Figure~\ref{fig:effect_of_K} shows the recall-QPS curves for $K = 1$ and $K = 100$ on the GIST dataset.
As $K$ increases, the routing becomes more challenging, yet \pname{} consistently accelerates the search by at least 30\% when targeting 80\% recall, demonstrating its robustness across different $K$.

\begin{figure}[t]
	\centering
	\subfigure[Probability density of angles on the DEEP dataset]{
		\begin{minipage}[b]{0.97\linewidth}
		\includegraphics[width=\linewidth]{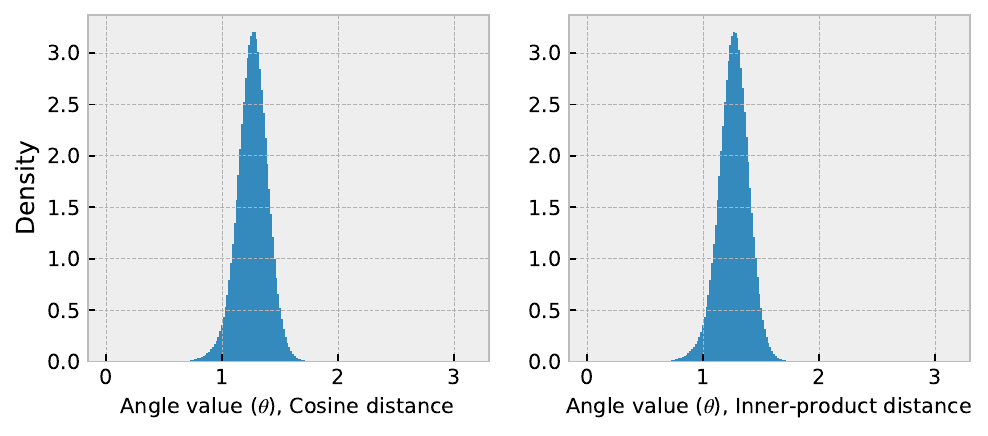}
		\end{minipage}
	}
	\subfigure[Recall-QPS curves on the DEEP dataset]{
		\begin{minipage}[b]{0.97\linewidth}
		\includegraphics[width=\linewidth]{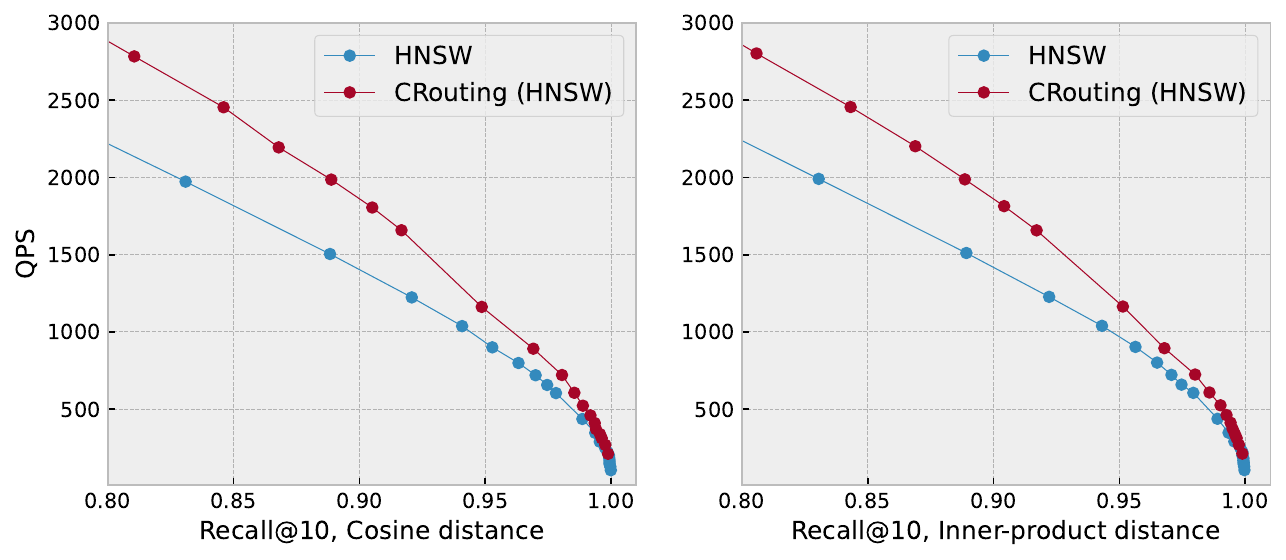}
		\end{minipage}
	}
	\caption{Generality across different distance metrics.}
	\label{fig:generality}
\end{figure}

\begin{figure}[t]
	\centering
	\subfigure[Recall-speedup curves on large datasets]{
		\begin{minipage}[b]{0.97\linewidth}
		\includegraphics[width=\linewidth]{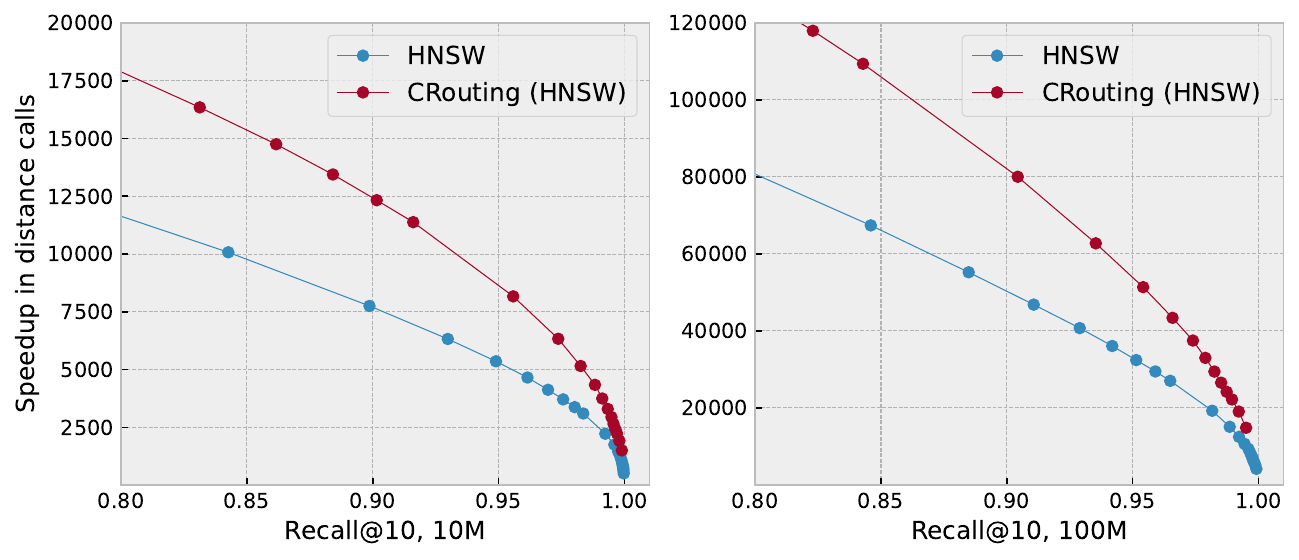}
		\end{minipage}
	}
	\subfigure[Recall-QPS curves on large datasets]{
		\begin{minipage}[b]{0.97\linewidth}
		\includegraphics[width=\linewidth]{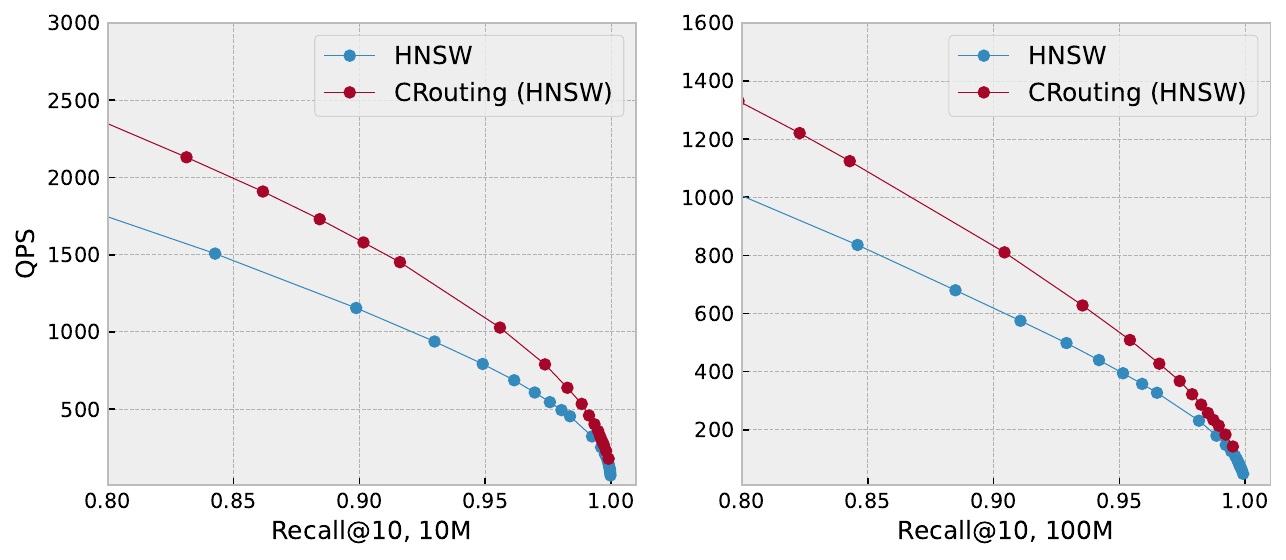}
		\end{minipage}
	}
	\caption{Scalability across different data volumes.}
	\label{fig:qps_bigann}
\end{figure}

\subsection{Generality and Scalability}
\label{eva:scalability and generality}
\textbf{Other distance metrics.}
We assess the generality of \pname{} for cosine and inner-product distance metrics. 
Figure~\ref{fig:generality}(a) shows that the angle distributions with other metrics are similar to that of the Euclidean distance. 
Using other distance metrics does not change the directional randomness between a node and its neighbors, and the angular distribution still exhibits a skewed characteristic.
Therefore, the pruning strategy of \pname{} remains effective.
As shown in Figure~\ref{fig:generality}(b), \pname{} achieves consistent improvements in QPS across different distance metrics.

\textbf{Search on large datasets.}
We evaluate the scalability of \pname{} in terms of data volume by sampling 10 million and 100 million vectors from the SIFT-1B dataset~\cite{sift-gist}. 
The SIFT-100M dataset represents the largest dataset that can be processed on our machine. Figure~\ref{fig:qps_bigann} shows the recall-speedup and recall-QPS curves. 
The results demonstrate that \pname{} consistently outperforms the baseline algorithms across all data volumes.


\begin{figure*}[t]
    \centering
    \includegraphics[width=\linewidth]{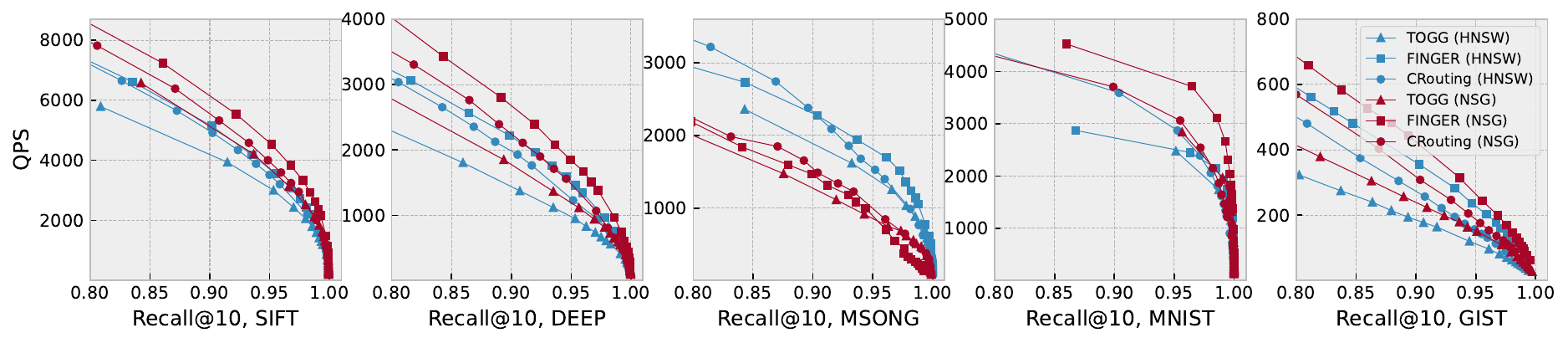}
    \caption{Recall-QPS curves for various routing strategies  (the top right is better).}
    \label{fig:recall-qps-routing}
\end{figure*}

\begin{table}[t]
    \small
    \caption{Construction time (sec) for various routing strategies.}
    \begin{tabular}{c|cccc}
    \hline
    Dataset & HNSW & +TOGG     & +FINGER   & +\pname{}    \\ \hline
    SIFT    & 677.1  & 707.8 (+5\%) & 771.7 (+14\%) & 685.9 (+1\%) \\
    DEEP   & 1163.1  & 1241.7 (+7\%) & 1397.1 (+20\%) & 1168.1 (+1\%) \\
    MSONG   & 1441.9   & 1508.6 (+5\%)  & 1702.6 (+18\%)  & 1461.8 (+1\%)  \\
    MNIST    & 46.1 & 49.2 (+7\%)& 81.6 (+78\%)& 46.5(+1\%)\\
    GIST   & 2881.8   & 3014.7 (+5\%)  & 3614.9 (+25\%)  & 2903.3 (+1\%)  \\ \hline
    Dataset & NSG     & +TOGG          & +FINGER        & +\pname{}        \\ \hline
    SIFT   & 163.1    & 193.6 (+19\%)  &   249.4 (+53\%)             & 169.1 (+4\%)      \\
    DEEP   &   265.2  & 350.8 (+32\%)  &   474.5 (+79\%)             &   272.3 (+3\%)             \\
    MSONG  &   317.8  & 389.9 (+23\%)  &    589.6 (+86\%)            &    325.7 (+2\%)            \\
    MNIST  &  13.8    & 16.6 (+20\%)   &   45.1 (+226\%)             &   14.3 (+4\%)             \\
    GIST   &    806.9 & 960.4 (+19\%)  &    1546.3 (+92\%)            &      814.4 (+1\%)          \\ \hline
    \end{tabular}
    \label{table:construction time}
\end{table}

\subsection{Comparison to Previous Routing Strategies}
\label{eva:comparison_to_other_strategies}

Figure~\ref{fig:recall-qps-routing} illustrates the recall-QPS curves for various routing strategies, 
with all competitors employing consistent index parameters.
Across all datasets, TOGG exhibits the poorest performance. This is primarily due to the use of the KD-tree for filtering out neighbors that differ in direction from the query node, which is often inaccurate in high-dimensional spaces, as confirmed by previous research~\cite{survey-vldb21}.
Both \pname{} and FINGER achieve higher QPS than TOGG. Although FINGER's QPS is approximately 10\% higher than \pname{}, it requires more construction time and incurs higher space costs for graph construction.

From the results presented in Table~\ref{table:construction time},
we observe that the construction time of CRouting increases by no more than 1\% on the HNSW algorithm and by no more than 4\% on the NSG algorithm.
Conversely, the construction time of FINGER increases by 14\% to 78\% on the HNSW algorithm, and increases by 53\% to 226\% on the NSG algorithm. 
FINGER requires more indexing time due to the need to construct an additional subspace for each node. 
This also explains why it has a higher QPS, as a substantial portion of the computational tasks during the search phase are offloaded to the construction process.
Table~\ref{table:index size} shows the results of index size.
The memory footprint of \pname{} increases by 2\% to 21\% due to the requirement to store the distances to neighboring nodes.
FINGER incurs pronounced space costs to store the information about projected vectors and LSH.
Specifically, the memory overhead of FINGER increases by 56\% to 295\% on the HNSW algorithm and by 82\% to 426\% on the NSG algorithm.

Construction time and space are critical performance metrics in vector databases.
Indexes for vector search are typically rebuilt weekly~\cite{JD-middleware18} to
balance low query latency, high accuracy, and the daily updates of billions of vectors. 
However, the process of graph reconstruction incurs significant resource overhead.  
For instance, building a global graph-based index for a 128 GB, 128-dimensional SIFT dataset with 1 billion vectors requires 1100 GB of memory over 2 days, or 5 days with 64 GB of memory and 32 vCPUs~\cite{diskann-nips19}. FINGER significantly increases both construction time and index size, making it particularly costly in such scenarios. \pname{} exhibits strong competitiveness in both construction and query efficiency for graph-based ANNS algorithms, making it a compelling choice for routing strategies.

\begin{table}[t]
    \small
    \caption{Index size (MB) for various routing strategies.}
    \begin{tabular}{c|cccc}
    \hline
    Dataset & HNSW & +TOGG     & +FINGER   & +\pname{}   \\ \hline
    SIFT    & 751.8  & 767.1 (+2\%)  & 2968.2 (+295\%) & 873.8 (+16\%)  \\
	DEEP    & 1240.1 & 1255.4 (+1\%) & 3456.5 (+178\%) & 1362.1 (+10\%)  \\ 
    MSONG   & 1851.3 & 1866.4 (+1\%) & 4050.5 (+119\%) & 1972.4 (+7\%) \\
    MNIST   & 195.2  & 196.2 (+1\%)  & 328.3 (+68\%)   & 202.5 (+4\%)  \\
    GIST    & 3925.6 & 3940.9 (+1\%) & 6142.1 (+56\%)  & 4047.7 (+3\%) \\ \hline
    Dataset & NSG    & +TOGG          & +FINGER        & +\pname{}      \\ \hline
    SIFT    &  620.8 & 636.2 (+3\%)  & 3264.5 (+426\%)   & 745.8 (+21\%)     \\
    DEEP    & 1142.1 & 1157.3 (+1\%) & 3785.7 (+232\%)   & 1299.8 (+14\%)             \\
    MSONG   & 1688.2 & 1703.3 (+1\%) & 4311.4 (+156\%)   & 1778.9 (+5\%)             \\
    MNIST   & 184.4  & 185.3 (+1\%)  & 343.1 (+86\%)    & 188.7 (+3\%)            \\
    GIST    & 3747.4 & 3762.7 (+1\%) & 6818.4 (+82\%)   & 3825.1 (+2\%)           \\ \hline
    \end{tabular}
    \label{table:index size}
\end{table}

\section{Related Work}
\label{sec:related work}
\textbf{Approximate Nearest Neighbor Search (ANNS).}
Existing ANNS algorithms can be categorized into four main types: (1) graph-based methods~\cite{nssg-tpami21,nsg-vldb19,nsw-is14,hnsw-tpami18,diskann-nips19,hcnng-pr19,survey-tkde19,fanng-cvpr16}, (2) quantization-based methods~\cite{rabitq-sigmod24,aq-cvpr14,imi-tpami14,optpq-cvpr13,itq-tpami12,accelerating-icml20,pq-tpami10,optpq-tpami13}, (3) tree-based methods~\cite{covertree-icml06,mtree-vldb97,randomprojectiontree-stoc08,flann-tpami14,revisitingkdtree-kdd19,hdindex-arxiv18,opKDtree-cvpr08,fastexact-siam13}, and (4) hashing-based methods~\cite{pstablelsh-scg04,dynamiclsh-sigmod12,qd-sigmod18,vhp-vldb20,srs-vldb14,lsbtree-tods10,pmlsh-vldb20,idec-vldb20,hashing4nns-vldb15}. Notably, graph-based approaches demonstrate superior performance for in-memory ANNS. In contrast, quantization-based methods excel in scenarios with limited memory resources.
Additionally, a substantial body of research has explored the application of machine learning techniques to accelerate searches~\cite{learingspace-iclr20,reinforcementrouting-tis23,learning2route1-icml19,learning2route2-icml20,optlearning-sigmod20}.
For further details, we direct readers to recent tutorials~\cite{newtrends-vldb21,tutorials-kdd21,survey-dataeng23}, as well as comprehensive reviews and benchmarks~\cite{survey-vldb21,survey-tkde19,annbenchmarks-is20,library-sisap13,review-da09}.

\textbf{Routing strategies.}
TOGG~\cite{togg-kbs21} and HCNNG~\cite{hcnng-pr19} utilize KD-trees to ascertain the direction of the query, effectively narrowing the search to vectors aligned with that specified direction. FINGER~\cite{finger-www23} leverages Locality Sensitive Hashing to estimate the distance of each neighbor from the query. Additionally, other machine learning-based optimizations~\cite{learning2route1-icml19,learning2route2-icml20,optlearning-sigmod20,bliss-sigmod22,flex-is24,learned-metric-is21} develop a routing function that employs supplementary representations to enhance routing efficiency from the starting node to the nearest neighbor. However, all these optimizations either sacrifice search accuracy or require extra computational resources, resulting in prolonged graph construction times.
In contrast, \pname{} strikes a favorable balance between search and construction efficiency.

\textbf{Other Optimizations.}
As the volume of vector data continues to increase, there has been a surge in interest around supporting incremental updates to vector indexes~\cite{spfresh-sosp23,freshdiskann-arxiv21,AdaIVF-arxiv24}, a crucial technique for enabling efficient and accurate for ANNS. Some studies~\cite{nhq-nips24,filterdiskann-www23} focus on developing efficient and robust frameworks for hybrid query processing, which integrates ANNS with attribute constraints.
Additionally, several efforts have optimized ANNS for newer hardware. For example, GPU-based ANN indexes such as SONG~\cite{song-icde20} and GGNN~\cite{ggnn-tbd22} have been proposed, achieving up to two orders of magnitude speedup over CPU-based methods. HM-ANN~\cite{hmann-nips20} redesigns the HNSW algorithm using Intel Optane persistent memory~\cite{optane-2019}. CXL-ANNS~\cite{cxlanns-atc23} decouples DRAM from the host via Compute Express Link~\cite{cxl-2020}, placing all essential datasets into its memory pool to handle billion-point graph-based ANNS.
FANNS~\cite{fanns-sc23} automatically co-designs hardware and algorithms on FPGAs based on user-defined recall requirements and hardware resource budgets.
These advanced optimizations are orthogonal to our work and can be incorporated in future research efforts.

\section{Conclusions}
\label{sec:conclusion}
We propose \pname{}, a novel routing strategy designed to enhance navigation in graph-based ANNS algorithms. 
\pname{} approximates the distance function by leveraging the angle distributions of high-dimensional vectors,
enabling the avoidance of unnecessary distance calculations.
It is designed as a plugin to optimize existing graph-based search with minimal code modifications.
Our experiments show that \pname{} reduces the number of distance computations by up to 41.5\% across two state-of-the-art algorithms, HNSW and NSG, while maintaining the same accuracy, thereby enhancing  the overall QPS by up to 1.48$\times$.


\bibliographystyle{ACM-Reference-Format}
\bibliography{ref}


\end{sloppypar}
\end{document}